# Switching intrinsic magnetic skyrmions with controllable magnetic anisotropy in van der Waals multiferroic heterostructures

Ze-quan Wang[1], Feng Xue[2], Liang Qiu[1], Zhe Wang[3], Ruqian Wu[3], Yusheng Hou[1,*]

[1] Guangdong Provincial Key Laboratory of Magnetoelectric Physics and Devices, Center for Neutron Science and Technology, School of Physics, Sun Yat-Sen University, Guangzhou, 510275, China

[2] Department of Physics, College of Physics & Optoelectronic Engineering, Jinan University, Guangzhou, Guangdong 510632, China

[3] Department of Physics and Astronomy, University of California, Irvine, CA 92697-4575, USA

**ABSTRACT**

Magnetic skyrmions, topologically nontrivial whirling spin textures at nanometer scales, have emerged as potential information carriers for spintronic devices. The ability to efficiently create and erase magnetic skyrmions is vital yet challenging for such applications. Based on first-principles studies, we find that switching between intrinsic magnetic skyrmion and high-temperature ferromagnetic states can be achieved in two-dimensional van der Waals (vdW) multiferroic heterostructure $CrSeI/In_2Te_3$ by reversing the ferroelectric polarization of $In_2Te_3$. The core mechanism of this switching is traced to the controllable magnetic anisotropy of CrSeI influenced by the ferroelectric polarization of $In_2Te_3$. We propose a useful descriptor linking the presence of magnetic skyrmions to magnetic parameters, and validate this connection through studies of a variety of similar vdW multiferroic heterostructures. Our work demonstrates that manipulating magnetic skyrmions via tunable magnetic anisotropies in vdW multiferroic heterostructures represents a highly promising and energy-efficient strategy for future development of spintronics.

**KEYWORDS: Intrinsic magnetic skyrmions, tunable magnetic anisotropy, van der Waals multiferroic heterostructures, two-dimensional ferromagnetism**

* Corresponding authors: houysh@mail.sysu.edu.cn

Magnetic skyrmions, characterized by their topologically nontrivial whirling spin textures, have been identified as promising high-density information carriers for the next-generation spintronic devices, such as skyrmion-based racetrack magnetic memory [1-5]. One crucial obstacle in advancing these technologies is the efficient creation and elimination of magnetic skyrmions through external stimuli. Currently, employing an electric field is recognized as a low-power approach for this purpose, supported by experimental observations in transition-metal multilayers [6-10]. Theoretical studies have unveiled that electric-field-induced variations in Heisenberg exchange interactions, Dzyaloshinskii–Moriya interactions (DMIs) and magnetic anisotropy are crucial for these experimental manipulations [11-13]. Given that the polarization characteristics of ferroelectric (FE) materials can provide a nonvolatile electric field affecting adjacent magnetic materials, it is natural to explore whether heterostructures combing FE and magnetic materials can serve as novel platforms for manipulating magnetic skyrmions. Indeed, this strategy has been substantiated in both experimental and theoretical studies involving multiferroic heterostructures of various oxides [14-17].

Since the first discovery in 2009 [18,19], magnetic skyrmions have been observed in a large array of magnetic materials, including chiral magnet MnSi bulks [18], B20-type $Fe_{0.5}Co_{0.5}Si$ thin films [20] and Fe monolayers (MLs) on the Ir surface [21]. Because of their synthetic feasibility and structural stability, two-dimensional (2D) van der Waals (vdW) magnetic materials and their heterostructures have emerged as a preferred option for hosting high-density magnetic skyrmions [22-37]. The inversion symmetry is typically broken at the interfaces in 2D magnetic vdW heterostructures, so strong DMIs are generated for the formation and stabilization of magnetic skyrmions. Many successful examples have been reported in previous theoretical and experimental studies [29, 31, 33-37]. Specifically, 2D magnetic Janus MLs, notable for their inherent loss of inversion symmetry and hence even stronger electric polarization, are poised to be optimal candidates for the realization of magnetic skyrmions [24-27]. This potential is particularly noticeable when they are combined with the recently discovered 2D FE materials [38-40], such as $In_2Se_3$ [41-43] and $In_2Te_3$ [44]. The vast array of 2D vdW materials, with their diverse properties and configurations, opens up new opportunities for controlling and utilizing magnetic skyrmions in a stable and efficient manner.

In this work, we perform systematical first-principles calculations to explore ways of manipulating magnetic skyrmions in vdW multiferroic heterostructures consisting of magnetic Janus ML Cr$YX$ ($Y$ = S, Se, Te; $X$ = Cl, Br, I) and 2D FE ML $In_2Y_3$ ($Y$ = Se, Te). Thanks to the FE proximity effect, magnetic anisotropies of Cr$YX$ can be largely modulated by reversing the FE polarization of $In_2Y_3$. Consequently, we find that CrSeI/$In_2Te_3$ can realize intrinsic Néel-type magnetic skyrmions when $In_2Te_3$ has a downward FE polarization. The radii of magnetic skyrmions may shrink to 2.7 nm from 8.7 nm by applying an external magnetic field up to 0.5 Tesla. When the FE polarization of $In_2Te_3$ is upward, CrSeI/$In_2Te_3$ adopts a desirable high-temperature (~250 K) ferromagnetic (FM) state, another important feature for the utilization of 2D magnetic materials. Finally, a useful descriptor featured by Heisenberg exchange interactions, DMIs and out-of-plane magnetic anisotropies is proposed to identify the appearance of magnetic skyrmions in these vdW multiferroic heterostructures. Our work demonstrates that engineering magnetic anisotropy of vdW multiferroic heterostructures by reversing the polarizations of 2D FE materials is a useful route towards electrical manipulations of magnetic skyrmions.

Since DMI plays a crucial role in the formation of magnetic skyrmions [45], considerable efforts have been dedicated to achieving a large DMI in magnets, typically by reducing symmetry and enhancing spin orbit coupling. Once DMIs become comparable to Heisenberg exchange interactions, magnetic skyrmions can be stabilized in the presence of uniaxial anisotropy [1, 45]. Given that the application of external magnetic fields represents the simplest approach to inducing uniaxial anisotropy, it is widely used in the generation of magnetic skyrmions. For instance, magnetic skyrmions were observed in the chiral itinerant-electron magnet MnSi when a small external magnetic field was applied [18]. Similarly, external magnetic fields can induce magnetic skyrmions in Janus magnet MnSTe [24] and CrTeI MLs [26], despite these materials lacking intrinsic magnetic skyrmions in the absence of such fields. The ability to tune out-of-plane uniaxial magnetic anisotropy through means like strain and electric field is a key area of interest as this tunability can be harnessed to control of magnetic skyrmions.

Owing to the intrinsic low symmetry and the presence of interfacial electric fields, magnetic anisotropy and DMI in Cr$YX$/In$_2Y_3$ systems are expected to be significantly altered by switching the polarization of FE substrates. Indeed, the polarization reversal induced magnetic reorientation has been demonstrated in various 2D vdW multiferroic heterostructures with CrI$_3$, Cr$_2$Ge$_2$Te$_6$ and CrTe$_2$ magnetic MLs [46-48]. We choose the recently proposed 2D magnetic Janus Cr$YX$ MLs [49] as our model systems because they have inherently broken inversion symmetry and heavy elements. Should it be confirmed that these materials possess large DMI and tunable magnetic anisotropy, it would imply that magnetic skyrmions in their multiferroic heterostructures can be activated or deactivated by reversing the polarization of FE substrates.

To this end, we construct a series of vdW multiferroic heterostructures consisting of magnetic Janus MLs of Cr$YX$ ($Y$ = S, Se, Te; $X$ = Cl, Br, I) and FE MLs of In$_2Y_3$ ($Y$ = Se, Te). Depending on the combinations of $Y$ and $X$, magnetic ground states of Cr$YX$ MLs are complex, including high-temperature FM orders, spin spiral orders and wormlike domains [26]. It is important to note that despite Cr$YX$ MLs having large DMIs according to theoretical calculations, intrinsic magnetic skyrmions have not yet been observed in these materials [26]. On the other hand, In$_2$Se$_3$ and In$_2$Te$_3$ MLs are 2D FE semiconductors [41, 44] and possess nonvolatile FE polarizations. Considering the mismatch between lattice constants of Cr$YX$ and In$_2Y_3$ MLs, we adopt $2 \times 2$ supercells of Cr$YX$ and $\sqrt{3} \times \sqrt{3}$ supercells of In$_2Y_3$ to build their heterostructures, and apply further slight stretch to the In$_2Y_3$ lattice for perfect match.

By calculating the binding energies of eight Cr$YX$/In$_2Y_3$ heterostructures with different stacking configurations (Table S1 and Figure S1), we find that $Y^{2-}$ ($Y$ = S, Se, Te) side of Cr$YX$ prefers to contact with In$_2Y_3$ ($Y$ = Se, Te), irrespective of the FE polarization directions of In$_2Y_3$. Without loss of generality, we take CrSeI/In$_2$Te$_3$ as an example to describe more stacking details. When In$_2$Te$_3$ has an upward FE polarization in CrSeI/In$_2$Te$_3$ (detonated as CrSeI/In$_2$Te$_3$(P↑)), atomic layers are stacked in a sequence of I-Cr-Se-Te-In-Te-In-Te, with the side and top views as shown in Figure 1a. In addition, CrSeI and $In_2Te_3$ are separated by interlayer distance of 3.38 Å. When $In_2Te_3$ has a downward FE polarization (detonated as CrSeI/$In_2Te_3$(P↓)), the geometry has no change excepts a slight expansion in the interfacial separation to 3.42 Å. A summary of the stacking details and magnetic ground states of Cr$YX$/$In_2Y_3$ is given in Table S1 and Figure S2.

To quantitatively describe the magnetic properties of Cr$YX$ in these heterostructures, we employ a spin Hamiltonian as follows:

$$H = J_1\sum_{<ij>}S_i \cdot S_j + J_2 \sum_{<<ij>>} S_i \cdot S_j + J_3 \sum_{<<<ij>>>} S_i \cdot S_j + D_{ij} \cdot \sum_{<ij>} S_i \times S_j - K\sum_i (S_i^z)^2 \quad (1).$$

In Eq. (1), $S_i$ ($S_j$) is the spin of the $i^{th}$ ($j^{th}$) Cr atom; $J_1$, $J_2$ and $J_3$ are the nearest neighbor (NN), second- and third-NN Heisenberg exchange parameters (Figure 1c), respectively; $D_{ij} = (D_x, D_y, D_z)$ is the NN DMI vector and $K$ is single ion anisotropy (SIA) parameter. Here, positive and negative $J_i$ mean antiferromagnetic (AFM) and FM Heisenberg exchange interactions. We employ the least-squares fitting technique [50] to determine Heisenberg exchange parameters and the four-state energy mapping method [51] to obtain the NN DMI vectors. For the convenience of following discussions, the NN DMI vector is reshaped as in-plane component, $D_{//} = \sqrt{D_x^2 + D_y^2}$ and out-of-plane one, $D_z$. Typically, the DMI is a short-range effect, predominately assigned as interactions among the nearest neighbors [12], as shown by our density functional theory (DFT) calculations of the NN and second-NN parameters presented in Table S2. Hence, only the NN DMI is considered in following analysis. It is worth noting that a positive (negative) $K$ means an out-of-plane (in-plane) magnetic anisotropy.

DFT calculated magnetic parameters of Cr$YX$/$In_2Y_3$ heterostructures are listed in Table 1 and Table S3. Compared with pristine Cr$YX$ MLs, the presence of $In_2Y_3$ leads to noticeable changes in all parameters of Cr$YX$/$In_2Y_3$. Nevertheless, the NN FM Heisenberg interactions in Cr$YX$/$In_2Y_3$ are still dominant and $K$ and $D$ are also somewhat strengthened. As our goal is to manipulate magnetic skyrmions by tuning the magnetic anisotropy in magnets with sizable DMIs, we focus on how the SIAs of Cr$YX$/$In_2Y_3$ are modulated when the polarization direction of $In_2Y_3$ is reversed. Note that the positive

SIA parameters of CrSI/In$_2$Te$_3$ and CrSeI/In$_2$Te$_3$ are remarkably enlarged when the FE polarization of In$_2$Te$_3$ is reversed from downward to upward. Compared with pristine CrSI and CrSeI MLs, both CrSI/In$_2$Te$_3$ and CrSeI/In$_2$Te$_3$ exhibit enhanced out-of-plane magnetic anisotropies. By contrast, the SIAs of the other six heterostructures are almost unaffected when the polarization direction of In$_2$Y$_3$ is reversed.

Based on Eq. (1) with DFT parameters, we perform MC simulations to determine the magnetic ground state of Cr$YX$/In$_2Y_3$. Specifically, magnetic skyrmions are identified by the topological charge $Q$ which reads [52]:

$$Q = \frac{1}{4\pi} \int \mathbf{m} \cdot \left( \frac{\partial \mathbf{m}}{\partial x} \times \frac{\partial \mathbf{m}}{\partial y} \right) dxdy \qquad (2).$$

In Eq. (2), **m** is a normalized magnetization vector; $x$ and $y$ are in-plane coordinates. On a discrete spin lattice, $Q$ is evaluated according to the Berg formula [53, 54]. Our MC simulations show that only CrSeI/In$_2$Te$_3$ displays a FE-polarization-induced switch of intrinsic magnetic skyrmions while other systems have either FM or spin spiral orders. When In$_2$Te$_3$ has an upward FE polarization, the magnetic ground state of CrSeI/In$_2$Te$_3$ is FM with a high Curie temperature ($T_C$) of 248 K (Figure S3). Excitingly, an intrinsic magnetic skyrmion with $Q = -1$ forms in CrSeI/In$_2$Te$_3$ when the FE polarization of In$_2$Te$_3$ is downward (Figure 2a). Additionally, this intrinsic magnetic skyrmion is the Néel-type and has a radius of 8.7 nm at 0.1 K. Because the manipulation of magnetic skyrmion is the central topic of this work, we will focus our discussions on CrSeI/In$_2$Te$_3$ hereafter.

To account for the thermal fluctuation induced destabilization of magnetic skyrmions, we study the temperature effect on the topological spin structures in CrSeI/In$_2$Te$_3$ under various external magnetic field ($B$). Particularly, we examine topological charge $Q$ under different $T$ and $B$. Here, the simulated ground states are identified by $Q$ and the real-space spin textures. Without an external magnetic field (i.e., $B = 0$), MC simulations reveal that the intrinsic Néel-type magnetic skyrmion in CrSeI/In$_2$Te$_3$ can be preserved below a critical temperature of $T_c^{SkX} = 156$ K (Figure 2b). When temperature is higher than $T_c^{SkX}$, the intrinsic Néel-type magnetic skyrmion is deformed (denoted as fluctuation-disorder phase) due to the strong thermal fluctuation and finally disappear at about 162 K. On the other hand, the radius of the intrinsic Néel-type magnetic skyrmion at 0.1 K shrinks to 2.7 nm from 8.7 nm when an external magnetic field of 0.5 Tesla is applied. However, it should be noted that a larger external magnetic field (i.e., $B > 0.5$ Tesla) may destroy magnetic skyrmions and change the system into the FM order.

It is exciting that the intrinsic Néel-type magnetic skyrmion in $CrSeI/In_2Te_3$ can survive with the temperature below 150 K and an external magnetic field smaller than 0.5 Tesla and, furthermore, the magnetic skyrmions can be switched on and off by reversing the FE polarization in $In_2Te_3$. To unveil its underlying mechanism, we inspect the magnetic parameters of $CrSeI/In_2Te_3(P\uparrow)$, $CrSeI/In_2Te_3(P\downarrow)$ and pristine CrSeI ML (Table 1), and noticed that the relative changes in $K$ are more obvious than those in $J$ and $D$. Taking the pristine CrSeI ML as a reference, relative changes in NN Heisenberg exchange parameters and DMIs are 5% for $CrSeI/In_2Te_3(P\uparrow)$ and 12% for $CrSeI/In_2Te_3(P\downarrow)$, respectively. However, SIA parameters are increased by about 158% for $CrSeI/In_2Te_3(P\uparrow)$ and 27% for $CrSeI/In_2Te_3(P\downarrow)$. Due to its strong out-of-plane magnetic anisotropy, $CrSeI/In_2Te_3(P\uparrow)$ prefers an out-of-plane FM order, as confirmed by MC simulations. By comparing magnetic parameters of $CrSeI/In_2Te_3$ and CrSeI, we perceive that the formation of magnetic skyrmions in $CrSeI/In_2Te_3$ is highly sensitive to the value of $K$.

To verify this hypothesis, we investigate the evolution of the magnetic ground state of $CrSeI/In_2Te_3$ by artificially varying $K$ in a large range in MC simulations. With $J$ and $D$ parameters fixed, magnetic skyrmion may also exist in $CrSeI/In_2Te_3(P\uparrow)$ when $K$ is tuned into an appropriate range (0.195-0.40 meV/Cr) (Figure 2c). Its magnetic ground state changes to either a spin spiral order or wormlike domain when $K$ is out of this range. As expected, strong in-plane or out-of-plane SIA leads to FM order. Interestingly, magnetic bimeron, a topological counterpart of magnetic skyrmion, can be realized when the in-plane magnetic anisotropy is tuned to a medium range (Figure 2c). Obviously, $CrSeI/In_2Te_3(P\uparrow)$ and $CrSeI/In_2Te_3(P\downarrow)$ have the same evolutions of magnetic states when $K$ is tuned, as shown in Figure 2c. Hence, their drastically different magnetic ground states almost solely stem from their "different" $K$ values.

Now, we study the electronic structure of CrSeI/In$_2$Te$_3$ under different polarization directions to understand how the FE polarization of In$_2$Te$_3$ affects SIA of CrSeI. As shown in Figure 3b-3c, freestanding CrSeI and In$_2$Te$_3$ MLs are semiconductors with band gaps of 1.39 eV and 0.61 eV, respectively. When they are put together, the heterostructure's band structure obviously depends on the polarization direction of the In$_2$Te$_3$ ML. Although both CrSeI/In$_2$Te$_3$(P↑) and CrSeI/In$_2$Te$_3$(P↓) have a type-II band alignment, the former has a much smaller band gap than the latter (Figure 3a, 3d). Correspondingly, the former has more apparent charge redistributions than the latter, as evidenced by their charge density differences (Figure S4). These results suggest that the upward FE polarization of In$_2$Te$_3$ has a stronger modulation on the electronic properties of CrSeI/In$_2$Te$_3$. By examining the orbital resolved magnetic anisotropic energy (MAE), we find that the positive MAEs of both CrSeI/In$_2$Te$_3$(P↑) and CrSeI/In$_2$Te$_3$(P↓) are mainly contributed by the hybridizations of 5$p$ orbitals of iodine atoms and 3$d$ orbitals of chromium atoms (Figure 3e-3f and Figure S5). Obviously, the MAE differences between CrSeI/In$_2$Te$_3$(P↑) and CrSeI/In$_2$Te$_3$(P↓) are mainly contributed by iodine atoms (Figure 3g). When the FE polarization of In$_2$Te$_3$ is reversed from upward to downward, the contributions from the hybridizations between $p_x$ and $p_y$ orbitals to the out-of-plane MAE are largely decreased. Overall, the tunable SIA in CrSeI/In$_2$Te$_3$ is closely related to contributions from iodine atoms, which are reactive to the switch of FE polarization in In$_2$Te$_3$.

Finally, we propose a helpful descriptor [55] to assist in screening vdW multiferroic heterostructures for the generation of magnetic skyrmions. As magnetic skyrmions are stabilized by the interplay between Heisenberg exchange interaction, DMI and out-of-plane magnetic anisotropy, the descriptor should include all of them. With this in mind, we introduce exchange stiffness $A$ [56], DMI coefficient $d$ and effective out-of-plane magnetic anisotropy $K_{eff}$ which are defined as follows:

$$A = \frac{1}{V_0} \sum_{ij} \frac{J_{ij}}{2} (a_{ij})^2 \quad (3),$$

$$d = \frac{n_1 D_{//} a_1}{2V_0} \quad (4),$$

$$K_{\text{eff}} = \frac{nK}{V_0} \quad (5).$$

In Eq. (3)-(5), $V_0$ is the volume of the unit cell; $a_{ij}$ is the distance between $i^{\text{th}}$ and $j^{\text{th}}$ Cr atoms; $n$ is the number of magnetic atoms in the unit cell; $n_1$ is coordination number of the NN Cr-Cr pairs; more details of $A$, $K_{\text{eff}}$ and $d$ are given in Supporting information. Based on $A$, $d$ and $K_{\text{eff}}$, we discover that a descriptor, $\eta = d/\sqrt{AK_{\text{eff}}}$, can effectively identify the potential magnetic skyrmion materials. According to the DFT data in this work, we find that magnetic skyrmions may emerge when $\eta$ is in the range of 0.69 to 0.98, such as in CrSeI/In$_2$Te$_3$(P↑) and CrSeI/In$_2$Te$_3$(P↓) as discussed for Fig. 3c. On the contrary, systems with an $\eta$ value beyond this specified range do not support the formation of stable magnetic skyrmions. This is not a surprise as a small $\eta$ means either strong FM Heisenberg exchange interactions between Cr atoms (i.e., large $A$) or large out-of-plane magnetic anisotropy energy (i.e., large $K$) and, accordingly, the FM order is energetically favored. On the opposite end, a large $\eta$ value corresponds to strong DMI, which tends to favor spin spiral or wormlike textures. As shown in Table S4, when Cr$YX$/In$_2Y_3$ heterostructures and pristine Cr$YX$ MLs have $\eta$ smaller than 0.67, they display a FM magnetic ground state. Conversely, the pristine CrSeI ML has $\eta = 1.05$ and it has a spin spiral order. Overall, the dimensionless descriptor $\eta$ appears to provide a reliable guidance for identifying 2D vdW multiferroic heterostructures for the generation and manipulation of magnetic skyrmions.

In summary, based on first-principles calculations and Monte Carlo simulations, we systematically studied magnetic properties of 2D vdW multiferroic heterostructure Cr$YX$/In$_2Y_3$. We found that intrinsic Néel-type magnetic skyrmions can be switched on and off in CrSeI/In$_2$Te$_3$ by reversing the FE polarization of In$_2$Te$_3$. This switching is attributed to the tunable magnetic anisotropy in CrSeI/In$_2$Te$_3$, via weak but meaningful hybridization between the semiconducting magnetic and FE materials. It is interesting that many such systems can host magnetic skyrmions when their single ion anisotropy is tuned into an appropriate window. From the experimental point of view, adjusting temperature and employing doping can also be utilized to slightly tweak the single ion anisotropy of Janus 2D magnets [57, 58], aiming to fine tune their MAE across or within such window to enhance the stability of skyrmions. In addition, we proposed a practical descriptor aimed at facilitating the identification of Néel-type magnetic skyrmions across a variety of 2D vdW multiferroic heterostructures. Our work demonstrates that tuning magnetic anisotropies in semiconducting vdW multiferroic heterostructures presents a promising approach for controlling the formation and elimination of magnetic skyrmions as required for practical applications.

**Methods.** *Density functional theory calculations.* Our first-principles calculations are based on the density-functional theory, using the functional at the level of the generalized gradient approximation (GGA) as implemented in Vienna *ab initio* simulation package (VASP) [59, 60]. We use a Hubbard parameter $U_{\text{eff}}$ = 3.0 eV to account for the strong correlation effects among the Cr 3$d$ electrons [26]. DFT-D3 method [61] is adopted to describe the vdW interactions between Cr$YX$ and In$_2Y_3$. Based on the magnetic parameters from DFT calculations, we perform Monte Carlo (MC) simulations with the Metropolis algorithm to explore the magnetic ground states of Cr$YX$/In$_2Y_3$.

■ **ASSOCIATED CONTENT**

**Supporting information**

The Supporting Information is available free of charge at XXX.

Detailed Information about DFT computational and MC simulation details, the different stacking configurations of Cr$YX$/In$_2Y_3$ heterostructures, Curie temperatures and magnetic ground states of Cr$YX$/In$_2Y_3$, charge density difference of CrSeI/In$_2$Te$_3$, details for obtaining orbital resolved MAE and orbital resolved MAE of CrSeI/In$_2$Te$_3$, magnetic parameters of Cr$YX$/In$_2Y_3$ and micromagnetic parameters of CrSeI/In$_2$Te$_3$.

■ **AUTHOR INFORMATION**

**Corresponding Authors**

**Yusheng Hou** - School of Physics, Guangdong Provincial Key Laboratory of Magnetoelectric Physics and Devices, Center for Neutron Science and Technology, Sun

Yat-Sen University, Guangzhou, 510275, China

Email: houysh@mail.sysu.edu.cn

**Authors**

**Ze-quan Wang** - School of Physics, Guangdong Provincial Key Laboratory of Magnetoelectric Physics and Devices, Center for Neutron Science and Technology, Sun Yat-Sen University, Guangzhou, 510275, China

**Feng Xue** - Department of Physics, Jinan University, Guangzhou, Guangdong 510632, China

**Liang Qiu** - School of Physics, Guangdong Provincial Key Laboratory of Magnetoelectric Physics and Devices, Center for Neutron Science and Technology, Sun Yat-Sen University, Guangzhou, 510275, China

**Zhe Wang** - Department of Physics and Astronomy, University of California, Irvine, CA 92697-4575, USA

**Ruqian Wu** - Department of Physics and Astronomy, University of California, Irvine, CA 92697-4575, USA

**Notes**

The authors declare no competing financial interest.

## ■ ACKNOLEGEMENTS

This work was supported by the National Key R&D Program of China (Grant No. 2022YFA1403301) and the National Natural Sciences Foundation of China (Grants No. 12104518, 92165204), GBABRF-2022A1515012643. The DFT calculations reported were performed on resources provided by the Guangdong Provincial Key Laboratory of Magnetoelectric Physics and Devices (No. 2022B1212010008) and Tianhe-II. Ruqian Wu acknowledges support from the USA-DOE, Office of Basic Energy Science (Grant No. DE-FG02-05ER46237). Feng Xue also acknowledges the support of China Postdoctoral Science Foundation (grant No. 2023M741388).

# Figures, tables and captions

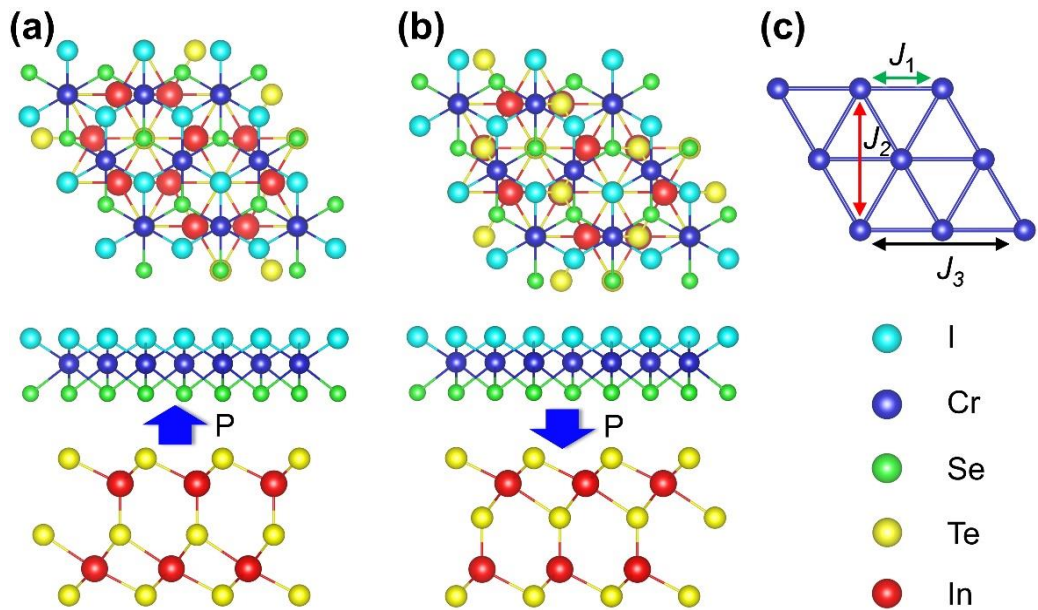

**Figure 1.** (a) Top and side views of the crystal structure of the most stable stacking configuration of CrSeI/In$_2$Te$_3$(P↑) heterostructure. The upward polarization of In$_2$Te$_3$ is indicated by the blue arrow. (b) same as (a) but for CrSeI/In$_2$Te$_3$(P↓). (c) The NN ($J_1$), second-NN ($J_2$) and third-NN ($J_3$) Heisenberg exchange paths are shown in the sublattice of Cr atoms.

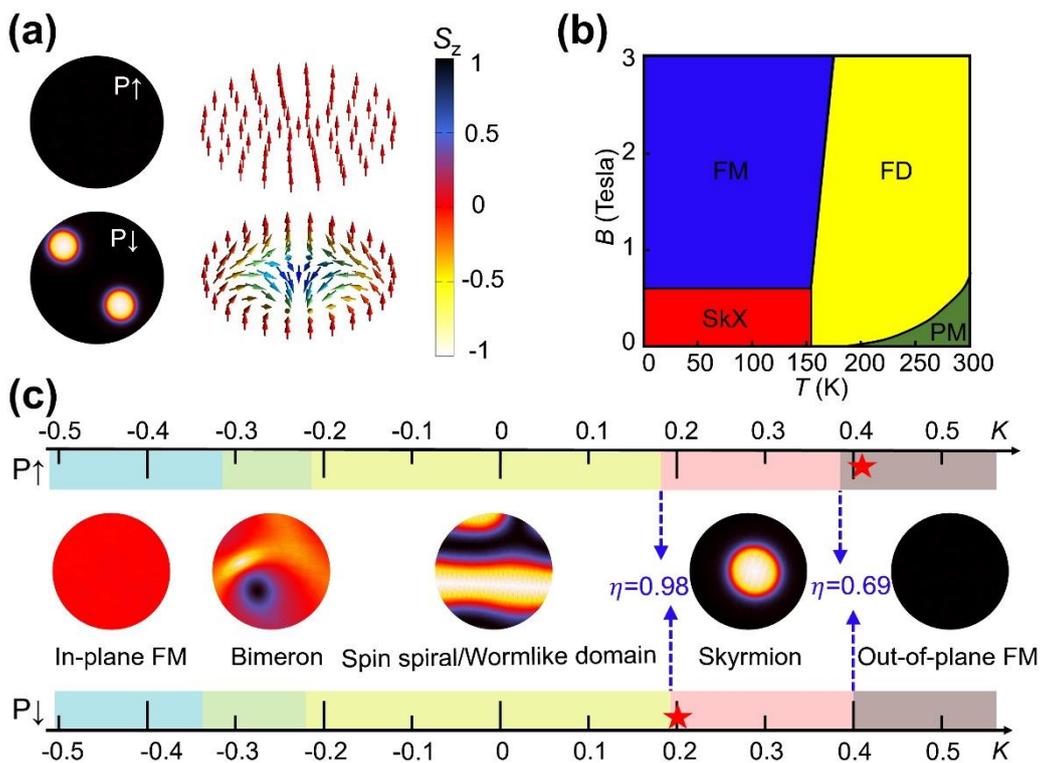

**Figure 2.** (a) Real-space spin textures of CrSeI/In$_2$Te$_3$ with upward (P↑) and downward (P↓) polarizations. (b) The phase diagram of CrSeI/In$_2$Te$_3$ in the *B-T* plane. SkX, FD and PM represent skyrmion, fluctuation-disorder and paramagnetic phases, respectively. (c) Evolutions of magnetic ground state with varied *K* (in unit of meV/Cr) in CrSeI/In$_2$Te$_3$(P↑) and CrSeI/In$_2$Te$_3$(P↓). The red stars highlight the DFT calculated *K* in CrSeI/In$_2$Te$_3$(P↑) and CrSeI/In$_2$Te$_3$(P↓). The color bar in (a) indicating out-of-plane spin components is applied to (c) as well.

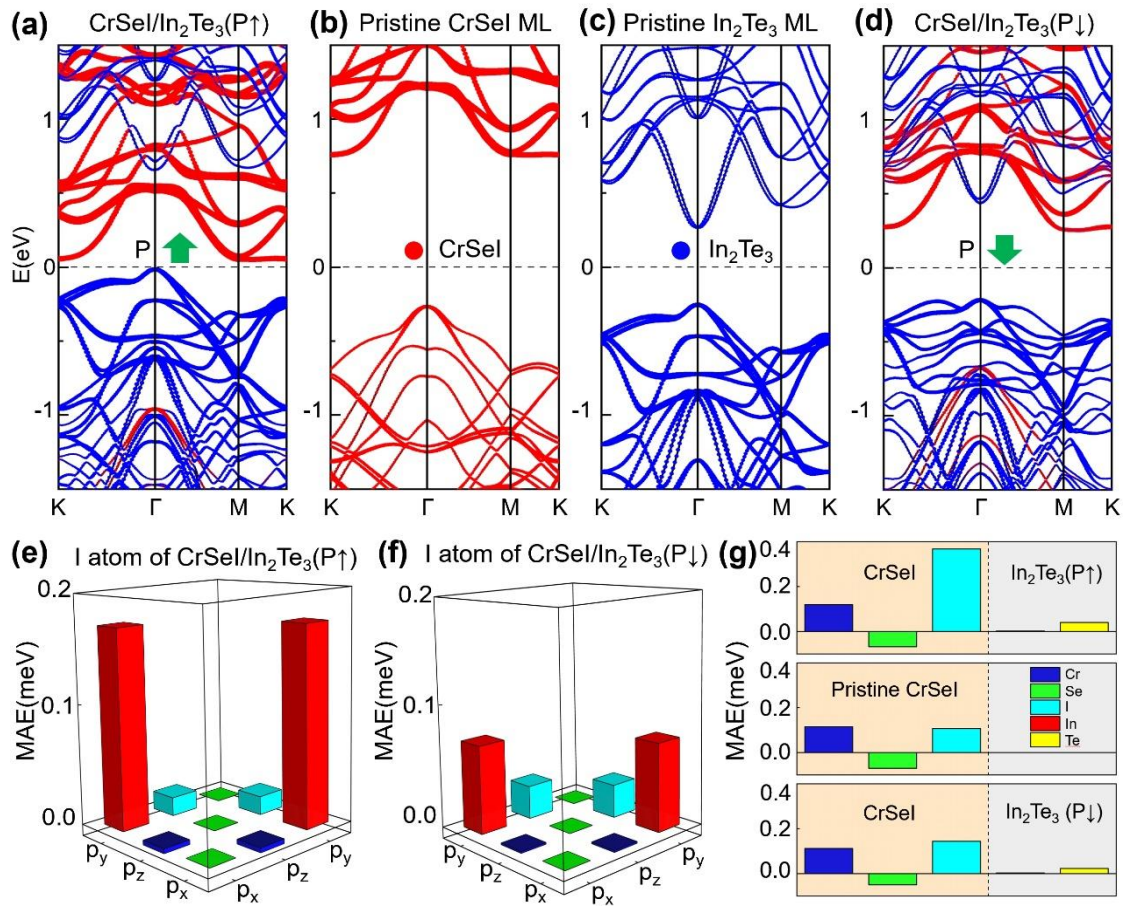

**Figure 3.** Band structures of (a) CrSeI/In$_2$Te$_3$(P↑), (b) pristine CrSeI ML, (c) pristine In$_2$Te$_3$ ML and (d) CrSeI/In$_2$Te$_3$(P↓). In (a) and (d), the bands from the states of CrSeI and In$_2$Te$_3$ are indicated by the red and blue dots, respectively. (e) and (f) show orbital resolved MAEs of iodine atoms in CrSeI/In$_2$Te$_3$(P↑) and CrSeI/In$_2$Te$_3$(P↓), respectively. (g) A summary of the atomic resolved MAE.

Table 1. Heisenberg exchange parameters $J_i$ ($i$ = 1, 2, 3), SIA parameter $K$ (meV/Cr), $D_\parallel$ and $D_z$ of the NN DMI vector, dimensionless parameter $\eta$ and magnetic ground states (MGS) of CrSeI/In$_2$Te$_3$ and the pristine CrSeI ML. Magnetic parameters $J_i$, $D_\parallel$ and $D_z$ are all in the unit of meV.

| System | $J_1$ | $J_2$ | $J_3$ | $K$ | $D_\parallel$ | $D_z$ | $\eta$ | MGS |
|---|---|---|---|---|---|---|---|---|
| CrSeI/In$_2$Te$_3$(P↑) | -17.625 | -1.057 | 0.194 | 0.409 | 1.504 | 0.325 | 0.66 | FM |
| Pristine CrSeI | -16.754 | -0.615 | 0.993 | 0.158 | 1.407 | 0.382 | 1.05 | spin spiral |
| CrSeI/In$_2$Te$_3$(P↓) | -18.893 | -1.149 | 0.272 | 0.201 | 1.612 | 0.177 | 0.97 | skyrmion |